\documentclass[a4paper,10pt,twoside]{cpc-hepnp}
\usepackage{upgreek,fancyhdr}
\usepackage{multicol}
\usepackage{graphicx}
\usepackage{booktabs}
\usepackage{amssymb,bm,mathrsfs,bbm,amscd}
\usepackage[tbtags]{amsmath}
\usepackage{lastpage}
\usepackage{amssymb}

\begin{document}
\fancyhead[c]{\small Submitted to Chinese Physics C}
\fancyfoot[C]{\small 010201-\thepage}


\title{Effects of Laser Pulse Heating of Copper Photocathodes on High-brightness Electron Beam Production at Blowout Regime\thanks{Supported by the National Natural Science Foundation of China (Grants No. 11375097)}}

\author{%
      Lian-Min Zheng$^{1,2,3}$%
\quad Ying-Chao Du$^{1,2}$%
\quad Chuan-Xiang Tang$^{1,2;1)}$\email{tang.xuh@tsinghua.edu.cn}%
\quad Wei Gai$^{1,2,3}$%
}
\maketitle

\address{%
$^1$ Accelerator Laboratory, Department of Engineering Physics, Tsinghua University, Beijing 100084, China\\
$^2$ Key Laboratory of Particle \& Radiation Imaging (Tsinghua University), Ministry of Education, Beijing 100084, China\\
$^3$ High Energy Physics Division, Argonne National Laboratory, Lemont, Illinois 60439, USA
}

\begin{abstract}
Producing high-brightness and high-charge ($>$100 pC) electron bunches at blowout regime requires ultrashort laser pulses with high fluence. The effects of laser pulse heating of the copper photocathode are analyzed in this paper.
The electron and lattice temperature is calculated using an improved two-temperature model, and an extended Dowell-Schmerge model is employed to calculate the thermal emittance and quantum efficiency. A time-dependent growth of the thermal emittance and the quantum efficiency is observed. For a fixed amount of charge, the projected thermal emittance increases with  decreasing laser radius, and this effect should be taken into account in  laser optimization at blowout regime. Moreover, laser damage threshold fluence is simulated, showing that the
maximum local fluence should be less than 40 $\mathrm{mJ/cm^2}$ to prevent damage to the cathode.
\end{abstract}



\footnotetext[0]{\hspace*{-3mm}\raisebox{0.3ex}{$\scriptstyle\copyright$}2013
Chinese Physical Society and the Institute of High Energy Physics
of the Chinese Academy of Sciences and the Institute
of Modern Physics of the Chinese Academy of Sciences and IOP Publishing Ltd}%

\begin{multicols}{2}

\section{Introduction}
\par In recent decades, high-brightness electron beams have been of central importance in many accelerator-based applications such as free electron laser (FEL) light sources~\cite{lab1}, MeV ultrafast electron diffraction~\cite{lab2}, and Thomson scattering X-ray sources~\cite{lab3}. One of the potential methods to achieve high brightness is to produce a beam with a three-dimensional uniformly filled ellipsoidal charge distribution. This charge distribution produces space-charge fields that have a linear dependence on position within the distribution, and gives a promising path to perfect emittance compensation performance. Several schemes have been proposed to generate the 3D uniform ellipsoidal distribution, including drive laser pulse shaping~\cite{lab4,lab5} and longitudinal space charge expansion~\cite{lab6}.
The latter technique uses an ultrashort laser illuminating a photocathode to create an initial charge distribution which is longitudinally thin and radially wide. This initial bunch, also referred to as the pancake bunch, will expand longitudinally under space charge forces to create a uniformly filled ellipsoid distribution. This space-charge-dominated expansion is usually referred to as the blowout regime. Producing 3D uniform ellipsoidal bunches at blowout regime is relatively simple to implement because ultrashort laser pulses are readily available and no more laser pulse shaping is required. Therefore, the blow-out regime scheme has attracted a lot of attention in the past decade and some experiments have been done to verify its performance in 3D uniform ellipsoidal bunch production~\cite{lab98,lab99}.

\par The blowout regime requires the characteristic image-charge field strength ($\sigma_0/\epsilon_0$) to be much smaller than the initial cathode field~\cite{lab6}, otherwise the image charge force will deform the charge distribution and dilute the transverse emittance~\cite{lab98}. Unfortunately, the accelerating field in emission phase at the photocathode is low ($\sim$60 MV/m) for convensional S-band RF guns, which significantly limits the achievable maximum charge density. This restriction will be strengthened for applications requiring high-charge electron bunches ($>$100 pC), like FEL applications. This is because for a limited charge density, the blowout regime requires large transverse laser spots at the photocathode with consequent large thermal transverse emittance~\cite{lab101}.
\par Recently, several photocathode RF guns with high accelerating gradients have been developed. For example, a 200 MV/m cathode accelerating field has been achieved in SLAC's X-band gun~\cite{lab100}, and even higher accelerating fields are expected in some novel photocathode RF gun designs, such as the cryogenic electron source~\cite{lab102}. The high accelerating field greatly eases the restriction of the charge density of electron bunches produced at blowout regime, and consequently, a small transverse emittance can be achieved more easily in high-gradient RF guns. Therefore, the blowout regime has become an important operation mode to produce high-brightness and high-charge bunches~\cite{lab100,lab102}.

\par The quantum efficiency is typically of the order of $10^{-5}$ for a conventional copper cathode. Producing high-brightness and high-charge electron bunches at blowout regime with such a low-quantum-efficiency cathode requires a laser beam with high energy, ultrashort laser pulses, and small transverse dimension, that is, the energy density of the laser beam is quite large. The heating effects of such a laser pulse on the electron bunch emission process should be analyzed carefully. Recently, a simulation of laser pulse heating of metallic photocathodes has been done~\cite{lab103}, showing a time-dependent growth of quantum efficiency and thermal emittance due to the ultrafast laser pulse heating effect. In this paper, we use this model to calculate the effects of laser pulse heating on high-brightness and high-charge electron bunch production at blowout regime. Given that the energy density of the laser pulse needed for the production of high-charge and ultrashort electron bunches is quite high, the lattice and electron temperatures will rise dramatically  to a significantly high level. Therefore, an improved two-temperature model is employed here to calculate the lattice and electron temperatures more accurately, considering the effects of lattice and electron temperatures on the reflectivity coefficient, the electron heat capacity, the electron thermal conductivity, and the electron-phonon coupling.


\section{Improved two-temperature model}
\par The ultrashort laser pulse heating of the metallic photocathode is a nonequilibrium energy transport process, which consists of two stages~\cite{lab8}. The first stage is photon energy absorption through  electron-photon interactions. It takes a few femtoseconds for electrons to reestablish the Fermi distribution. The second stage is  lattice temperature growth through electron-phonon interactions, which takes a few or tens of picoseconds. Consequently, there exists a state in which the electrons have reached an equilibrium state with very high electron temperature $T_e$ (thousands of kelvin), while the lattice temperature $T_l$ is still low (close to room temperature). This process is successfully described by the well-known two temperature model. In this paper an improved two-temperature model~\cite{lab9,lab10} is employed, which considers the effects of lattice and electron temperatures on the reflectivity coefficient, the electron heat capacity, the electron thermal conductivity, and the electron-phonon coupling over a wide range.
\par Generally, the diameter of laser beam used to illuminate the cathode is much larger than the sum of the optical penetration depth and the electron ballistic range. Therefore, a one-dimensional model is accurate enough to be used here, as given below:
\begin{equation}\label{eq4}
\begin{array}{l}
{C_e}({T_e})\frac{{\partial {T_e}}}{{\partial t}} = \frac{{\partial}}{{\partial z}} \left[ {{K_e}({T_e},{T_l})\frac{{\partial {T_e}}}{{\partial z}}} \right] - G({T_e} - {T_l}) + Q(z,t)\\
{C_l}({T_l})\frac{{\partial {T_e}}}{{\partial t}} = G({T_e} - {T_l})
\end{array}
\end{equation}
\par where $C_e$ is the electron heat capacity, $C_l$ is the lattice heat capacity, $K_e$ is the electron thermal conductivity, $G$ is the electron-phonon coupling constant, and $Q(z,t)$ represents the laser source term.
\par The parameters for $C_e=\gamma T_e$ with $\gamma=96.6$ $\mathrm{J/cm^3K^2}$ and $G=10^{17}$ $ \mathrm{W/m^3K}$ are widely used in many works~\cite{lab10,lab103}. These estimations are only valid, however, at low electron temperature ($0<T_e<3000$ K)\cite{lab12}. As the electron temperature becomes significantly high ($>$3000 K for copper), numerous d-band electrons are excited, leading to a significant increase of the electron heat capacity and electron-phonon coupling. Hence, Zhibin Lin's modified values (see Fig.~3 (c)(d) in Ref.~\cite{lab12}) are used in our model to estimate $C_e$ and $G$ in a wider electron temperature range.
\par The electron thermal conductivity can be described by~\cite{lab10}
\begin{equation}
K_e(T_e,T_l)=K_0\frac{BT_e}{AT_e^2+BT_l}
\end{equation}
\par where $K_0=400$ $\mathrm{W/mK}$ is the electron thermal conductivity at room temperature, $A=1.75\times10^7$ $\mathrm{K^{-2}\cdot S^{-1}}$ is the coefficient of e-e collision frequency, and $B=1.98\times10^{11}$ $\mathrm{K^{-1}\cdot S^{-1}}$ is the coefficient of e-ph collision frequency.
\par The lattice heat capacity is estimated by a Debye model.
\begin{equation}
C_l=9N{k_B}{\left( {\frac{T_l}{{{\theta _D}}}} \right)^3}\int_0^{{{{\theta _D}} \mathord{\left/
 {\vphantom {{{\theta _D}} T_l}} \right.
 \kern-\nulldelimiterspace} T_l}} {dx\;{x^4}} \frac{{{e^x}}}{{{{({e^x} - 1)}^2}}}
\end{equation}
\par where $N$ is the atomic number density, $k_B$ is the Boltzmann constant, and $\theta_D$ is the  Debye temperature ($\theta_D=343.5 K $ for copper).
\par A laser beam with Gaussian longitudinal profile and uniform radial profile is used in our model. The heat source term $Q(z,t)$ can be expressed as
\begin{equation}
Q(z,t) = \frac{[{1 - R({T_l})}]F_0}{{\sqrt {2\pi } {\sigma _t}{d_p}}}\exp [ - \frac{{{{(t - {t_0})}^2}}}{{2{\sigma _t}^2}} - \frac{z}{{{d_p}}}]
\end{equation}

\par where $F_0$ is the laser fluence, $\sigma_t$ is the rms laser pulse width with $t_0=4\sigma_t$, and ${d_p} = {1 \mathord{\left/{\vphantom {1 {\alpha  + {\lambda _{ball}}}}} \right.
 \kern-\nulldelimiterspace} {\alpha  + {\lambda _{ball}}}}$ is the effective penetration depth, which is the sum of the optical penetration depth ${1 \mathord{\left/
 {\vphantom {1 \alpha }} \right.
 \kern-\nulldelimiterspace} \alpha }$
 and the ballistic electron penetration depth $\lambda_{ball}$. Here we ignore the effect of temperature on the optical penetration depth and assume ${1 \mathord{\left/
 {\vphantom {1 \alpha }} \right.
 \kern-\nulldelimiterspace} \alpha }$ to be a constant, ${1 \mathord{\left/
 {\vphantom {1 \alpha }} \right.
 \kern-\nulldelimiterspace} \alpha }=12$ $\mathrm{nm}$\cite{lab103}. The measured value of ballistic electron penetration depth has shown a significant difference in different groups~\cite{lab15,lab16}. This value may be affected by specific characteristics of the sample, such as the crystallinity. We used the latest measured value $\lambda_{ball}=15$ $\mathrm{nm}$~\cite{lab16} in our model.
 \par R is the reflection coefficient and it has a significant dependence on the cathode temperature. Especially when the lattice temperature gets close to the melting point (1357.77 K for copper), the reflection coefficient will become very small. Therefore, the dependence of R on the lattice temperature $T_l$ should be analyzed carefully. The reflectivity coefficient is determined as follows:
\begin{equation}\label{eq8}
R=\frac{(n-1)^2+k^2}{(n+1)^2+k^2}
\end{equation}
 where $n+ik$ is the complex refractive index. The relation between the complex refractive index and the complex dielectric function is given as $n+ik=\sqrt{\epsilon}=\sqrt{\epsilon_1+i\epsilon_2}$, i.e., $n=\sqrt{(\epsilon_1+\sqrt{\epsilon_1^2+\epsilon_2^2})/2}$, $k=\sqrt{(-\epsilon_1+\sqrt{\epsilon_1^2+\epsilon_2^2})/2}$.
\par The complex dielectric function $\epsilon$ of metals is determined by the Drude model\cite{lab17} for free electrons, which is written as
\begin{equation}\label{eq9}
\epsilon=1-\frac{\omega_p^2}{\omega_L(\omega_L-i\nu)}
\end{equation}
where $\omega_p=\sqrt{\frac{n_ee^2}{m_e\epsilon_0}}$ is the plasma frequency (with $n_e$ the electron number density), $\omega_L$ is the laser frequency, and $\nu$ is the electron collision frequency. The collision frequency is usually calculated by Spitzer's formula~\cite{lab18} in many works about the metal ablation~\cite{lab19}, in which the free electrons are treated as an ideal gas. But this assumption is valid only if the electron temperature is much higher than the Fermi energy (Fermi temperature, $8\times10^4$ K for copper). Our concern in this study is only with the lattice temperature below the melting point and the electron temperature will not be so high. A formula to describe the dependence of the collision frequency on the lattice temperature above room temperature and below the melting point can be found in Ref. \cite{lab20}:
\begin{equation}\label{eq10}
\nu\approx2k_s\frac{e_1^2k_BT_l}{\hbar^2v_F}
\end{equation}
Here, $e_1$ is electron charge in electrostatic units, $\hbar$ is the reduced Planck constant, $v_F=\hbar(3\pi^2n_e)^{1/3}/m_e$ is the
Fermi velocity, and $k_s$ is a numerical constant.

\par
\par Considering a 266-nm laser beam illuminating a copper cathode, the dependence of reflectivity coefficient $R$ on the lattice temperature is shown in Fig.~\ref{R_Tl}. Here we assume $k_s=85$ in Eq.~(\ref{eq10}) to make the reflectivity coefficient at room temperature calculated by Eq.~(\ref{eq8})-(\ref{eq10}) consistent with the nominal value ($R=0.34$~\cite{lab103}).

\begin{center}
\includegraphics[width=7cm]{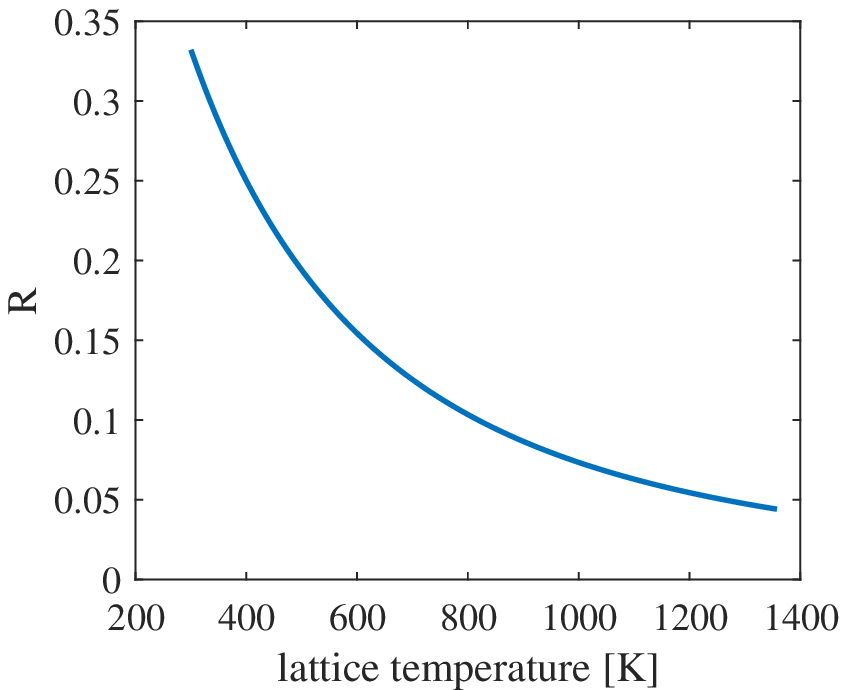}
\figcaption{\label{R_Tl} Dependence of reflectivity coefficient $R$ on the lattice temperature (from room temperature to melting temperature) when a 266-nm laser beam illuminates a copper cathode. }
\end{center}

\par For equation set (1), the initial condition can be written as
\begin{equation}\label{eq12}
T_e(z,0)=T_l(z,0)=T_0
\end{equation}
and the boundary conditions can be expressed as
\begin{equation}\label{eq13}
\begin{array}{l}
{{\partial {T_e}} \over {\partial z}}{|_{z = 0}} = {{\partial {T_e}} \over {\partial z}}{|_{z = d}} = 0  \\
   {{\partial {T_l}} \over {\partial z}}{|_{z = 0}} = {{\partial {T_l}} \over {\partial z}}{|_{z = d}} = 0
\end{array}
\end{equation}
Here, $T_0$ is the initial temperature, and $d=600$ nm is the boundary for numerical calculation, which is much larger than the effective penetration depth.

\par Equation set~(\ref{eq4}) and Eqs.~(\ref{eq12}), (\ref{eq13}) are numerically solved by a pdepe~\cite{lab21} function in MATLAB. When a 266-nm laser beam with rms pulse width $\sigma_t$ 30 fs and laser fluence $F_0$ 10 $\mathrm{mJ/cm^2}$ illuminates the cathode, the space- and time-dependence of the electron and lattice temperature in the copper cathode are shown in Fig.~\ref{fig2}. For a fixed time, the electron and lattice temperature decrease along with depth, and the maximum temperature appears at the cathode surface.


\par Figure~\ref{fig3} shows the temporal evolution of the electron and lattice temperature at the cathode surface with different laser fluences. The electrons in the copper cathode quickly absorb the energy of the laser beam through electron-photon collisions and the electron temperature increases to the maximum rapidly (in tens of femtoseconds). The temperature of the lattices rises much more slowly than that of the electrons because the energy transfer through electron-phonon collisions is much more difficult than through electron-photon collisions. It will take a few picoseconds for the electrons and lattices to reach a thermal equilibrium.

\end{multicols}
\ruleup
\begin{center}
\includegraphics[width=12cm]{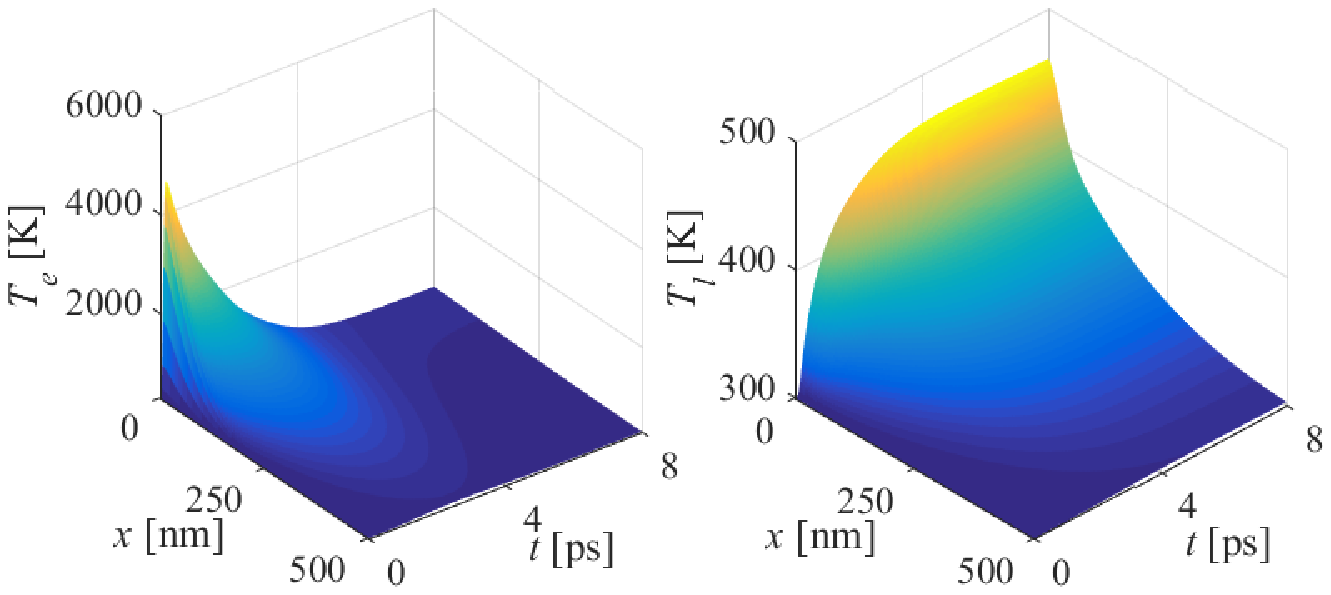}
\figcaption{\label{fig2} (color online) Electron temperature distribution (left) and lattice temperature distribution (right) at different times and positions for a copper cathode illuminated by a 266-nm laser beam with pulse width 30 fs, and laser fluence 10 $\mathrm{mJ/cm^2}$.}
\end{center}
\ruledown
\begin{multicols}{2}

\begin{center}
\includegraphics[width=7.5cm]{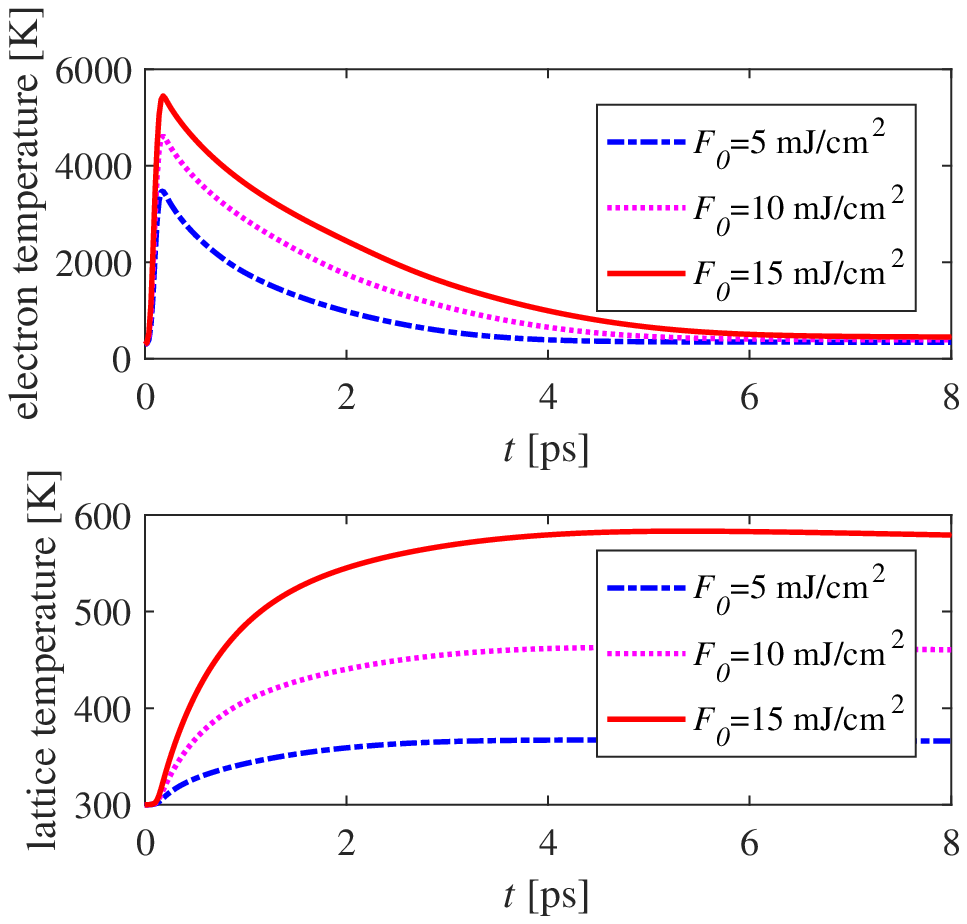}
\figcaption{\label{fig3} Temporal evolution of electron temperature (top) and lattice temperature (bottom) for a copper cathode illuminated by a 30-fs, 266-nm laser pulse at 5 $\mathrm{mJ/cm^2}$, 10 $\mathrm{mJ/cm^2}$, 15 $\mathrm{mJ/cm^2}$, respectively.}
\end{center}

\par Electron bunch emission at blowout regime requires that the acceleration-field-induced velocity-position
correlations associated with the finite duration $\tau_l$ of the photoemission process should be negligible, i.e.,
the laser pulse duration should be far less than the length of the electron bunches after full expansion. This requirement can be expressed as~\cite{lab6}
\begin{equation}\label{eq99}
  \tau_l\ll\frac{mc}{eE_0}
\end{equation}
where $\tau_l$ is the laser pulse duration, and $E_0$ is the cathode accelerating field in the emission phase.
\par For a high cathode accelerating field, the laser pulse duration is required to be very short. For example, a laser pulse with 60-fs FWHM length~\cite{lab100} was used in SLAC's X-band gun with a cathode accelerating field of 200 MV/m. In Fig.~\ref{fig4} we scan the rms laser pulse width $\sigma_t$ around the length of the laser used in SLAC's X-band gun, and get the dependence of the lattice and electron temperature on the rms laser pulse width. The maximum temperature of lattice and electrons changes little in the range of 15 fs $\le\sigma_t\le$ 80 fs. As a result, we almost cannot reduce the pulse heating effect by increasing the laser length at blowout regime in a high-accelerating-field gun.

\begin{center}
\includegraphics[width=7.5cm]{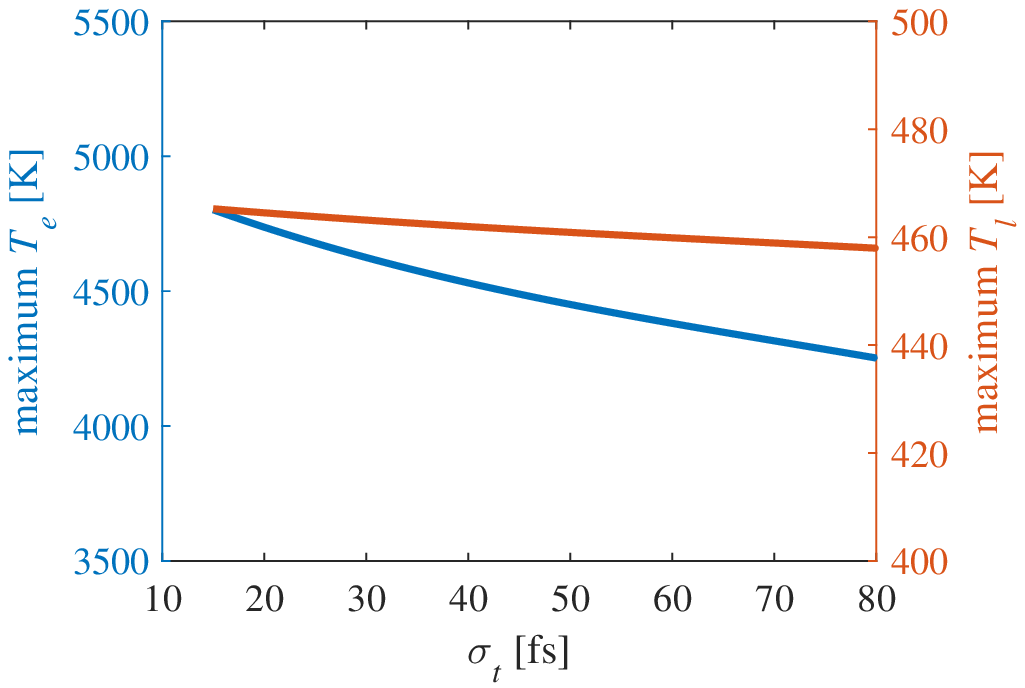}
\figcaption{\label{fig4} (color online) Maximum temperature of lattice and electrons as a function of rms laser pulse width. The laser fluence is fixed at 10 $\mathrm{mJ/cm^2}$. }
\end{center}

\section{Quantum efficiency and thermal emittance}
\par An extended Dowell-Schmerge model, which includes the effect of the finite temperature on the Fermi-Dirac distribution of the electrons in metals, is used here to calculate the quantum efficiency and the thermal emittance~\cite{lab23}. In this model the quantum efficiency can be expressed as
\begin{equation}\label{eq14}
  QE = {S_{12}}{{{{{\mathop{\rm Li}\nolimits} }_2}( - {\rm exp}[{{{E_{ex}}} \mathord{\left/
 {\vphantom {{{E_{ex}}} {k_B{T_e}}}} \right.
 \kern-\nulldelimiterspace} {k_B{T_e}}}])} \over {{{{\mathop{\rm Li}\nolimits} }_2}( - {\rm exp}[{{{E_f}} \mathord{\left/
 {\vphantom {{{E_f}} {k_B{T_e}}}} \right.
 \kern-\nulldelimiterspace} {k_B{T_e}}}])}}
\end{equation}
and the thermal emittance can be expressed as
\begin{equation}\label{eq15}
{\varepsilon _{n,x}} = {\sigma _{l,x}}\sqrt {{{k_BT_e} \over {{m_e}{c^2}}}} \sqrt {{{{{{\mathop{\rm Li}\nolimits} }_3}( - {\rm exp}[{{{E_{ex}}} \mathord{\left/
 {\vphantom {{{E_{ex}}} {k_B{T_e}}}} \right.
 \kern-\nulldelimiterspace} {k_B{T_e}}}])} \over {{{{\mathop{\rm Li}\nolimits} }_2}( - {\rm exp}[{{{E_{ex}}} \mathord{\left/
 {\vphantom {{{E_{ex}}} {k_B{T_e}}}} \right.
 \kern-\nulldelimiterspace} {k_B{T_e}}}])}}}
\end{equation}

where $T_e$ represents the electron temperature, and $E_{ex}=h\nu-\phi_{eff}$ represents the excess energy of electron emission, $E_f$ is the Fermi energy (7 eV for copper), $S_{12}$ is a constant related to the penetration probability for electrons through the cathode surface, and ${\rm{L}}{{\rm{i}}_n}$ is a poly-logarithm function defined
as
\begin{equation}\label{eq16}
{\rm{L}}{{\rm{i}}_n}(z) = {{{{( - 1)}^{n - 1}}} \over {(n - 2)!}}\int_0^1 {{1 \over t}} \log^{n - 2}{(t)}\log (1 - zt)dt
\end{equation}
\par The photon absorption and electron emission occur mainly in a few optical penetration depths, which is small enough to assume that the electron temperature is uniform in this depth, and the electron temperature at the cathode surface is used in Eq.~(\ref{eq14}) and Eq.~(\ref{eq15}) to calculate the quantum efficiency and the thermal emittance. For instance, Fig.~\ref{fig5} shows the quantum efficiency and thermal emittance as a function of time for a 266-nm laser pulse with rms pulse width 30 fs and fluence 10 $\mathrm{mJ/cm^2}$ illuminating the cathode.

\begin{center}
\includegraphics[width=7cm]{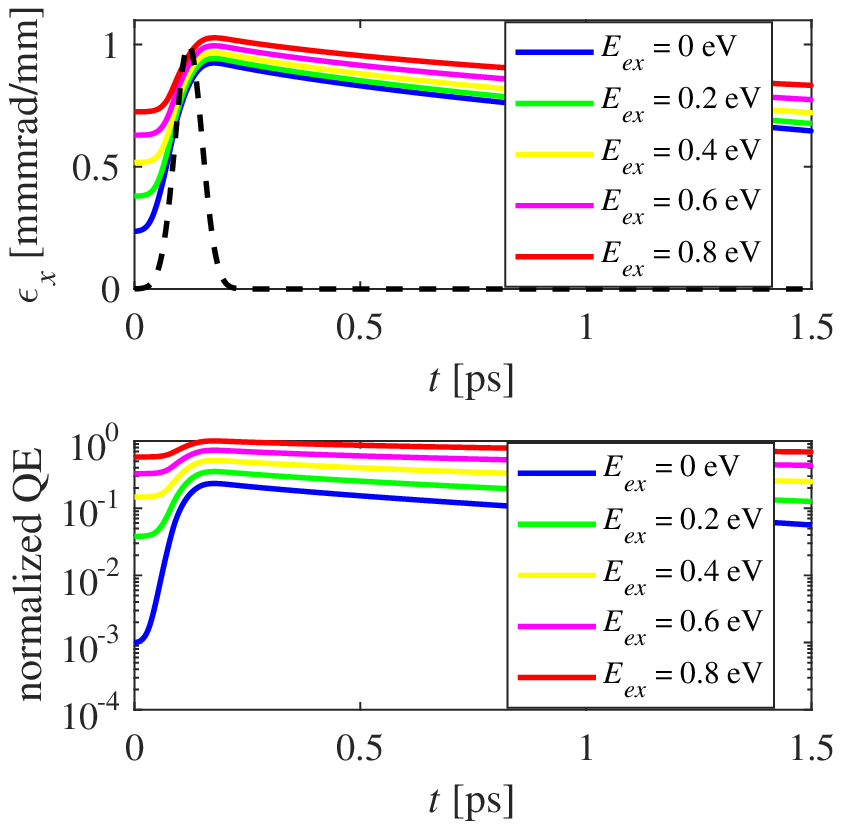}
\figcaption{\label{fig5} (color online) Thermal emittance (top) and quantum efficiency (bottom) as a function of time for different excess energies $E_{ex}$. The laser profile is plotted by the black dashed line. The rms laser pulse width $\sigma_t$ is 30 fs and the laser fluence $F_0$ is 10 $\mathrm{mJ/cm^2}$. The QE is normalized to the maximum at $E_{ex} = 0.8$ $\mathrm{eV}$.}
\end{center}

\par As shown in Fig.~\ref{fig5}, with the laser incidence and the corresponding electron temperature growth, the quantum efficiency and the thermal emittance increase rapidly. This effect is more pronounced in the case of low excess energy. When $t=$0 and the electron temperature is at room temperature, the thermal emittance can be approximated as $\sigma_x\sqrt{\frac{E_{ex}}{3m_ec^2}}$, which is determined by the excess energy $E_{ex}$. When the electron temperature becomes significantly high, the excess energy can be ignored (set $E_{ex}$=0), and the thermal emittance can be approximated as $\sigma_x\sqrt{\frac{k_BT_e}{m_ec^2}}$, which is determined only by the electron temperature. In Fig.~\ref{fig5} we find that the maxima of thermal emittance become similar for different excess energies, which means that the thermal emittance mainly comes from  electron thermalization, instead of excess energy $E_{ex}$, when the electron temperature becomes significantly high.

\par The excess energy $E_{ex}$ depends on the laser wavelength, the cathode work function, the accelerating field, and the field enhancement factor. An excess energy of 0.6 eV is assumed in the following simulations, equivalent to a thermal emittance of 0.63 mmmrad/mm at room temperature. Besides, the quantum efficiency at room temperature is assumed to be $2\times10^{-5}$~\cite{lab100} and the rms laser width is fixed at 30 fs in the following calculation. In Fig.~\ref{fig6} we plot the longitudinal profile of the electron bunch by multiplying the laser longitudinal profile by the time-varying quantum efficiency. As shown in Fig.~\ref{fig6}, although the quantum efficiency varies with time, the longitudinal profile of the electron bunch does not change much compared to the laser profile. The longitudinal profile of the electron bunch can still be approximated as a Gaussian distribution.

\begin{center}
\includegraphics[width=7.5cm]{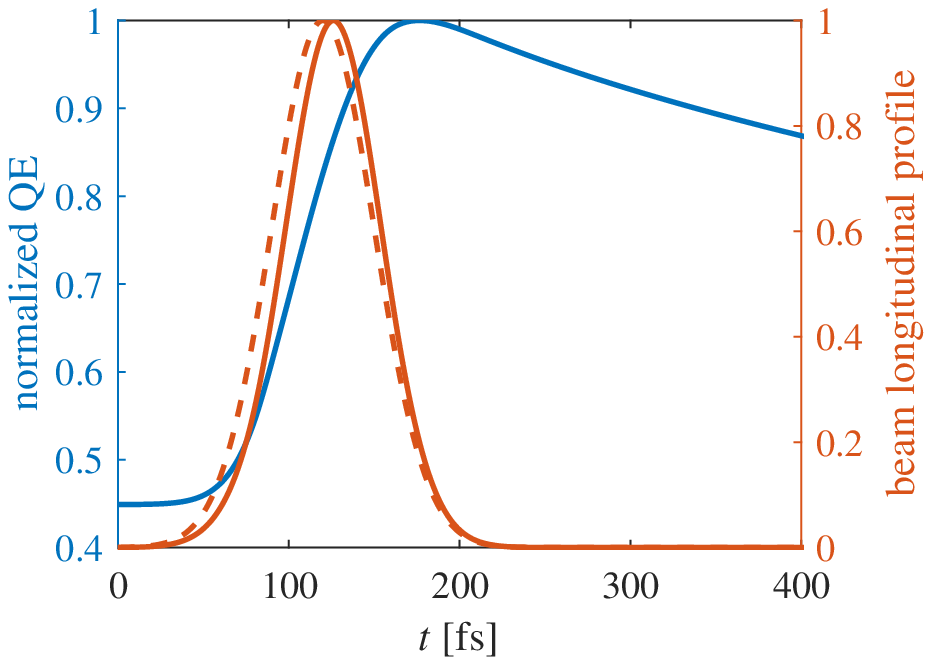}
\figcaption{\label{fig6} (color online) The blue line is the normalized QE as a function of time. The red dashed line is the laser longitudinal profile, and the red solid line is the electron bunch longitudinal profile. The rms laser pulse width $\sigma_t$ is 30 fs and the laser fluence $F_0$ is 10 $\mathrm{mJ/cm^2}$. The excess energy is assumed to be 0.6 eV.}
\end{center}
\par The charge density of the electron bunch as a function of the laser fluence is plotted in Fig.~\ref{fig7}. For a fixed QE at room temperature ($2\times10^{-5}$), the charge density should be proportional to the laser fluence. However, a time-varying QE has to be taken into account when the laser fluence is large. In this case, the average QE will be larger than $2\times10^{-5}$ and the curve of the charge density varying with the laser fluence becomes nonlinear.

\begin{center}
\includegraphics[width=7cm]{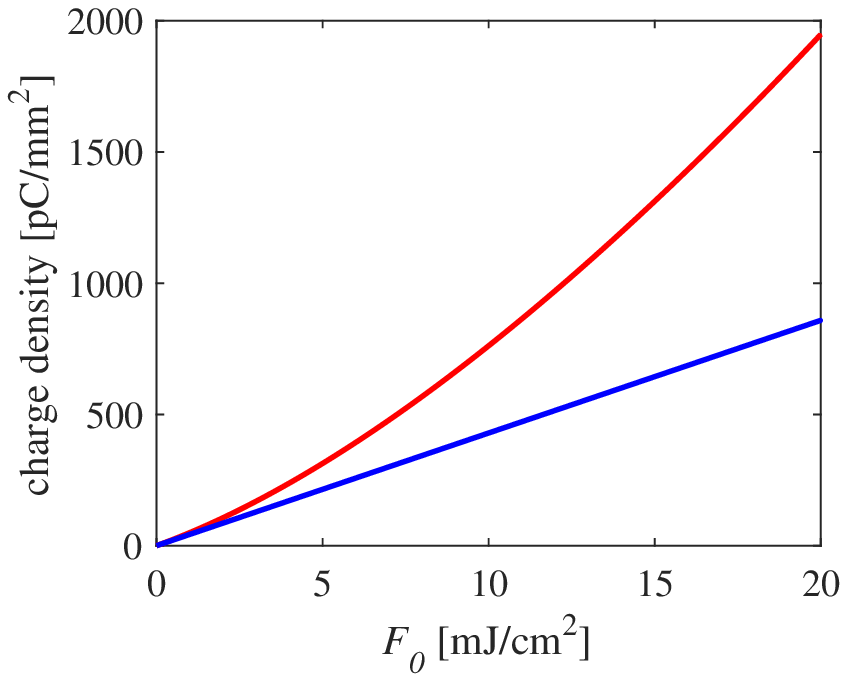}
\figcaption{\label{fig7} (color online) The charge density of the electron bunch as a function of the laser fluence. The blue line indicates that the quantum efficiency is fixed at $2\times10^{-5}$. The red line takes the time-varying QE into account, and the initial QE at room temperature is $2\times10^{-5}$.}
\end{center}

\par The laser's longitudinal and transverse dimensions should be carefully optimized to produce high-brightness and high-charge electron bunches at blowout regime. For instance, the laser pulse length should be short based on the accelerating field limitation (see Eq.~\ref{eq99}). Moreover, the laser radius needs to be carefully weighed, since a large laser radius results in a large thermal emittance, whereas a small laser radius results in bunch spatial distortion induced by the image charge.
\par In addition to the above factors, the impact of laser pulse heating also needs to be considered in optimization of the laser's longitudinal and transverse dimensions.
From the point of view of laser pulse heating, there is no significant limit to the laser pulse length because the electron and lattice temperatures change little with the laser pulse length at blowout regime (see Fig.~\ref{fig4}). However, the laser radius optimization under the effect of laser pulse heating should be analyzed carefully. Here we take an example of a 150-pC electron bunch produced at blowout regime, with a uniform laser transverse distribution assumed. As shown in the upper plot of Fig.~\ref{fig8}, a small laser radius implies a large charge density for a fixed amount of charge, and correspondingly, a large laser fluence is required. Here we select two laser radii 0.25 mm (a) and 0.6 mm (b) as an example. The electron bunch distributions for these two radii are shown in the bottom plot, in which the position distribution is obtained by the process shown in Fig.~\ref{fig6}. The thermal emittance is marked with a color map and two bunches are drawn under the same axis in order to share the same color map. The thermal emittance increases with time for both  laser radii, but the thermal emittance increases more for the smaller radius because a larger laser fluence is required for a smaller radius.
\begin{center}
\includegraphics[width=7.5cm]{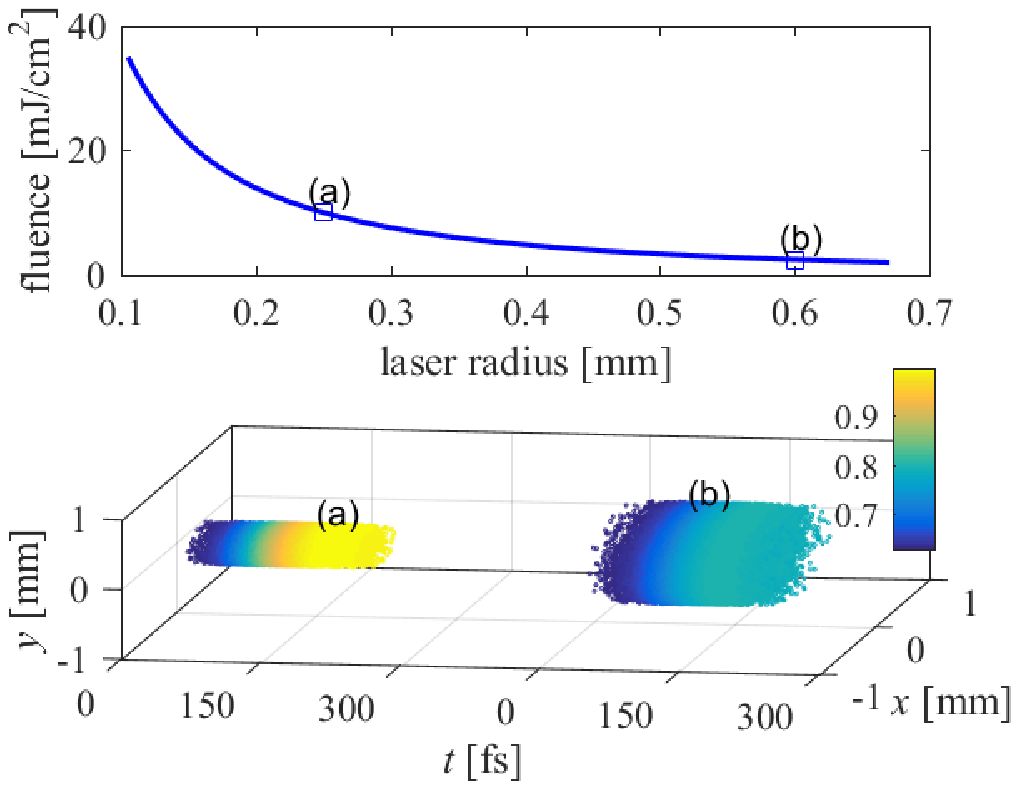}
\figcaption{\label{fig8} (color online) Upper plot: the laser fluence required to produce a 150-pC electron bunch for different laser radii. Here we select two laser radii 0.25 mm (a) and 0.6 mm (b), and plot the thermal emittance distribution in the bunch for these two points, as shown in the bottom plot. Here the unit of thermal emittance is mmmrad/mm.}
\end{center}

\par  A projected thermal emittance is calculated to characterize the thermal emittance of the entire bunch, which is written as 
\begin{equation}\label{eq4}
{\varepsilon _{proj}} = \frac{{\int {\varepsilon (\zeta )} I(\zeta )d\zeta }}{{\int {I(\zeta )} d\zeta }}
\end{equation}
\par where $\zeta$ is the relative position within the bunch, and $I(\zeta )$ represents the longitudinal current profile of the bunch.
\par The projected thermal emittance dependence on the laser radius is shown in Fig.~\ref{fig9}. 
\begin{center}
\includegraphics[width=7.5cm]{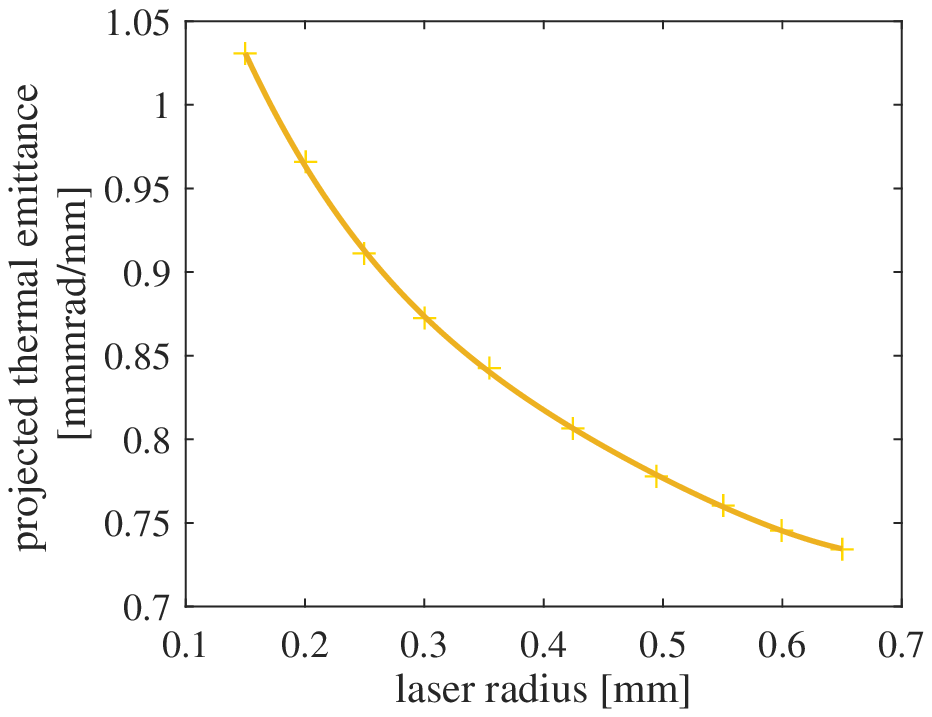}
\figcaption{\label{fig9} The projected thermal emittance as a function of the laser radius.}
\end{center}

The projected thermal emittance increases with decreasing laser radius, and this effect should be taken into account in laser optimization at blowout regime.

\section{Damage Threshold Fluence}
\par A laser beam with high energy density is used to produce high-charge electron bunches at blowout regime, which has the potential to damage the cathode. The damage threshold laser fluence should therefore be analyzed carefully. A 266-nm laser beam with rms pulse width of 30 fs is used to calculate the lattice temperature, and the lattice temperature as a function of the laser fluence is shown in Fig.~\ref{fig10}. 

\begin{center}
\includegraphics[width=7.5cm]{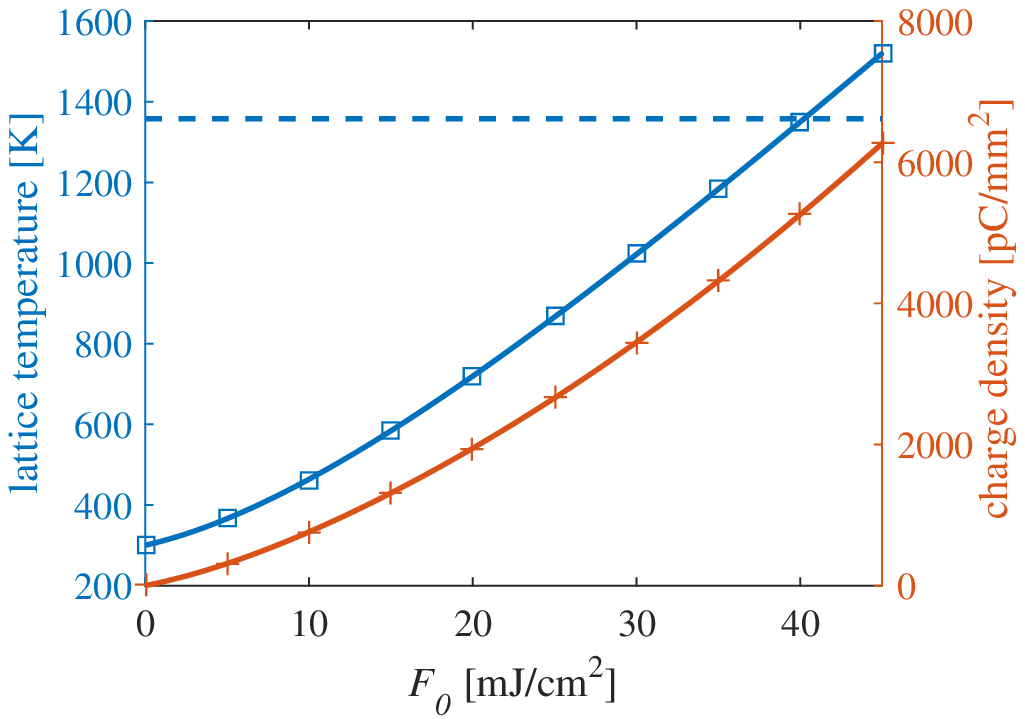}
\figcaption{\label{fig10} (color online) Blue solid line: the lattice temperature increases with the increase of the laser fluence. Blue dashed line: melting temperature of the cathode, 1357.77 K for copper. Red line: the charge density as a function of the laser fluence.}
\end{center}

\par By assuming the damage starts when the maximum lattice temperature reaches the melting temperature, 1357.77 K for copper, our model gives 40 $\mathrm{mJ/cm^2}$ for the threshold fluence. The charge density of the extracted electron bunch for the corresponding laser fluence is also plotted in Fig.~\ref{fig10}. According to our simulation, the charge density of the extracted bunch will be as high as 5200 $\mathrm{pC/mm^2}$ when cathode damage occurs. Such a high charge density requires at least 600 MV/m of accelerating field for the electron bunch to be extracted from the cathode according to Gauss's law. Such a high accelerating field is very difficult to achieve in the present gun design. Therefore, the concern about cathode damage is not necessary for an ideal transverse uniform laser beam. However, the transverse distribution of the laser beam generated in the laboratory is usually nonuniform. In this case, the charge density will be determined by the average laser fluence, whereas the damage threshold will be calculated by the maximum local fluence. That is to say, the maximum local fluence, rather than the average fluence, should be less than 40 $\mathrm{mJ/cm^2}$. For example, in Ref.~\cite{lab24} two ablation points were found on a cathode illuminated by a laser beam with an average fluence of 10 $\mathrm{mJ/cm^2}$, which is much smaller than the damage threshold calculated here, and that is because of the nonuniform energy distribution of the laser beam. Therefore, the transverse distribution of the laser beam should be as uniform as possible and the maximum local fluence should be less than 40 $\mathrm{mJ/cm^2}$ to prevent damage to the cathode.

\section{Conclusion}
\par Producing high-brightness and high-charge ($>$100 pC) electron bunches at blowout regime requires ultrashort laser pulse with high fluence. Given that the energy density of the laser pulses is quite high, the effects of laser pulse heating of the copper photocathode on the electron bunch emission process were analyzed in this paper. The electron and lattice temperature distributions were calculated using an improved two-temperature model. In this model the effects of lattice and electron temperature on the reflectivity coefficient, the electron heat capacity, the electron thermal conductivity, and the electron-phonon coupling were considered over a wide range. An extended Dowell-Schmerge model was employed to calculate the thermal emittance and quantum efficiency. A time-dependent growth in thermal emittance and quantum efficiency was observed. Since the average QE increases with the increase of laser fluence, a nonlinear dependence of the charge density on the laser fluence is observed in our simulation. For a fixed amount of charge, the projected thermal emittance increases with  decreasing laser radius, and this effect should be taken into account in  laser optimization at blowout regime. Moreover, laser damage threshold fluence was simulated, showing that the maximum local fluence should be less than 40 $\mathrm{mJ/cm^2}$ to prevent damage to the cathode.

\end{multicols}

\vspace{-1mm}
\centerline{\rule{80mm}{0.1pt}}
\vspace{2mm}

\begin{multicols}{2}

\end{multicols}

\clearpage

\begin{thebibliography}{90}

\vspace{3mm}

\bibitem{lab1}P. Emma, R. Akre, J. Arthur et al, Nat. Photonics, \textbf{4}(9): 641-647 (2010)

\bibitem{lab2} R. K. Li, and P. Musumeci, Phys. Rev. Applied, \textbf{2}(2): 024003 (2014)

\bibitem{lab3} Y. Du, L. Yan, J. Hua et al, Rev. Sci. Instrum., \textbf{84}(5): 053301 (2013)

\bibitem{lab4} C. Limborg-Deprey, and P. R. Bolton, Nucl. Instrum. Methods Phys. Res., Sect. A \textbf{557}: 106 (2006)

\bibitem{lab5} Y. Li, and J. W. Lewellen, Phys. Rev. Lett. \textbf{100}: 074801 (2008)

\bibitem{lab6} O. J. Luiten, S. B. van der Geer, M. J. de Loos et al, Phys. Rev. Lett., \textbf{93}(9): 094802 (2004)

\bibitem{lab98} P. Musumeci, J. T. Moody, R. J. England et al, Phys. Rev. Lett., \textbf{100}(24): 244801 (2008)

\bibitem{lab99} P. Piot, Y.-E Sun, T. J. Maxwell et al, Phys. Rev. ST AB, \textbf{16}(1): 010102 (2013)

\bibitem{lab101} Y. Li, arXiv:0809.1582

\bibitem{lab100} C. Limborg-Deprey, C. Adolphsen, D. McCormick et al, Phys. Rev. ST AB, \textbf{19}(5): 053401 (2016)

\bibitem{lab102} J.B. Rosenzweig, A. Cahill, V. Dolgashev et al, arXiv:1603.01657

\bibitem{lab103} J. Maxson, P. Musumeci, L. Cultrera et al, arXiv:1606.05629v2

\bibitem{lab8} T. Q. Qiu, and C. L. Tien, Int. J. Heat Mass Transfer, \textbf{35}(3): 719-726 (1992)

\bibitem{lab9} L. Jiang, and H. L. Tsai, J. Heat Transfer, \textbf{127}(10): 1167-1173 (2005)

\bibitem{lab10} R. Fang, D. Zhang, H. Wei et al, Laser Part. Beams, \textbf{28}(01): 157-164 (2010)

\bibitem{lab12} Z. Lin, L. V. Zhigilei, V. Celli, Phys. Rev. B, \textbf{77}(7): 075133 (2008)

\bibitem{lab15} J. Hohlfeld, S. S. Wellershoff, J. Gudde et al, Chem. Phys., \textbf{251}(1): 237-258 (2000)

\bibitem{lab16} J. Byskov-Nielsen, J. M. Savolainen, M. S. Christensen et al, Appl. Phys. A, \textbf{103}(2): 447-453 (2011)

\bibitem{lab17} J. B. Lee, K. G. Kang, and S. H. Lee, Mater. trans., \textbf{52}(3): 547-553 (2011)

\bibitem{lab18} L. Spitzer, \emph{Physics of fully ionized gases}, First edition (New York: Interscience Publishers, 1956), p. 78

\bibitem{lab19} L. Jiang, and H. L. Tsai, Int. J. Heat Mass Transfer, \textbf{48}(3): 487-499 (2005)

\bibitem{lab20} K. Eidmann, J. Meyer-ter-Vehn, T. Schlegel et al, Phys. Rev. E, \textbf{62}(1): 1202 (2000)

\bibitem{lab21} https://www.mathworks.com/help/matlab/ref/pdepe.html, retrieved 13th November 2016

\bibitem{lab23} T. Vecchione, D. Dowell, W. Wan et al, Quantum Efficiency and Transverse Momentum from Metals, in \emph{Proceedings of FEL2013}, (Geneva, Switzerland: JACoW, 2013), p. 424.

\bibitem{lab24} J. Shao, H. Chen, Y. Du et al, Phys. Rev. ST AB, \textbf{17}(7): 072002 (2014)

\end{thebibliography}
\end{document}